\documentclass{article}
\usepackage[T1]{fontenc} 
\usepackage[utf8]{inputenc} 
\usepackage{ismir,amsmath,cite,url}
\usepackage{graphicx}
\usepackage{color}
\usepackage{booktabs}
\usepackage{multirow}
\usepackage{caption}
\usepackage{subcaption}
\usepackage{amssymb}
\usepackage{adjustbox}
\usepackage{enumitem}
\usepackage{titlesec}
\usepackage{acronym}

\title{A Contrastive Self-Supervised Learning scheme for beat tracking amenable to few-shot learning }

\multauthor
{Antonin Gagneré \hspace{1cm} Slim Essid \hspace{1cm} Geoffroy Peeters} {LTCI - Télécom Paris, Institut Polytechnique de Paris, France\\
{\tt\small antonin.gagnere@telecom-paris.fr}
}

\def\authorname{A. Gagnere, S. Essid, and G. Peeters}

\acrodef{ssl}[SSL]{Self-Supervised Learning}
\acrodef{mir}[MIR]{Music Information Retrieval}
\acrodef{plp}[PLP]{Predominant Local Pulse}
\acrodef{osf}[OSF]{Onset Strength Function}
\acrodef{tcn}[TCN]{Temporal Convolutional Network}
\acrodef{dbn}[DBN]{Dynamic Bayesian Network}
\acrodef{fsl}[FSL]{Few-Shot Learning}

\begin{document}

\maketitle
\begin{abstract}
  In this paper, we propose a novel Self-Supervised-Learning scheme to train rhythm analysis systems and instantiate it for few-shot beat tracking.
  Taking inspiration from the Contrastive Predictive Coding paradigm, we propose to train a Log-Mel-Spectrogram-Transformer-encoder to contrast observations at times separated by hypothesized beat intervals from those that are not.
  We do this without the knowledge of ground-truth tempo or beat positions, as we rely on the local maxima of a Predominant Local Pulse function, considered as a proxy for Tatum positions, to define candidate anchors, candidate positives (located at a distance of a power of two from the anchor) and negatives (remaining time positions).
  We show that a model pre-trained using this approach on the unlabeled FMA, MTT and MTG-Jamendo datasets can successfully be fine-tuned in the few-shot regime, \textit{i.e.} with just a few annotated examples to get a competitive beat-tracking performance.
\end{abstract}
\section{Introduction}\label{sec:introduction}

Beat-tracking, \textit{i.e.} locating the times in a musical audio signal where beats are perceived or notated in the corresponding score, is still one of the most challenging subjects in the \ac{mir} research field.
This is owing to the large use of the beat information in many applications and to the complexity of the task: beats belong to a hierarchy/tree of rhythmic accentuations (hence entailing ambiguities), arise both from perceptual and cognitive cues. It, therefore, requires knowledge of the cultural specificities of the studied music.

To alleviate these issues, data-driven systems purely rely on training data composed of music tracks that have been annotated (supposedly) by experts.
However, this labeling process remains costly and as a consequence, the amount of data annotated into beats (at most a few thousands tracks) remains extremely low in \ac{mir}, as compared to other research fields (speech or computer vision).
For this reason, developing approaches that allow training beat-tracking systems without annotated data, a.k.a. \acf{ssl}, is important. 
This is the goal of this paper.

By alleviating the need of large annotated datasets, \ac{ssl}, has recently gained significant attention in the field of machine learning.
The goal is to learn meaningful representations of the input data without the need for human annotations.
To do so, the target outputs are directly inferred from the dataset itself, and often referred to as "pretext-task labels".
Such supervision can be obtained by masking some part of the input and asking the model to predict it \cite{bert, mae,bao2022beit,huang2022masked} or to generate two views of the same input and force a model to learn similar representations for the two views \cite{simclr, byol,simsiam,moco}.
Another popular \ac{ssl} approach is contrastive learning \cite{oord2019representation, wav2vec2} where one trains a network to predict whether two inputs are from the same class (or not) by forcing their trained embeddings to be more or less close from each other.
Usually, upon pre-training completion, the model is fine-tuned in a supervised fashion for one or more downstream tasks, where the data is smaller in size.
Our contributions are the following:
\begin{itemize}[topsep=0pt, partopsep=0pt, itemsep=0pt, parsep=0pt]
  \item We propose a novel contrastive \ac{ssl} scheme producing representations which are useful for automatic rhythm analysis tasks, in particular the beat-tracking task. Its key component is  the pretext-task design exploiting \acf{plp} local maxima to effectively sample anchor, positive, and negative time-steps for our contrastive loss function.
  \item We show that the pre-trained model can be fine-tuned in a few-shot learning setting to get competitive beat-tracking results. Moreover, we show that our approach yields, in most cases, at least better performance than Zero-Note Samba (ZeroNS)~\cite{desblancs2023zero}, which is, to the best of our knowledge, the only alternative \ac{ssl} approach to this problem to date.
  \item Furthermore we show that our model outperforms ZeroNS in a cross-dataset generalization setting.
  \item Finally we compare our model to the state of the art in a 8-fold cross-validation setting and show that it is competitive.
\end{itemize}

\textbf{Paper organization.}
The paper is organized as follows.
In section \ref{sec:related_work} we present works related to our proposal.
In section \ref{sec:method} we present our proposed contrastive \ac{ssl} training strategy.
Finally in section \ref{sec:expe} we present the results of the different experiments we performed.
To facilitate reproducibility, we make our code available.\footnote{https://github.com/antoningagnere/ssl\_beat}
\section{Related Work}\label{sec:related_work}

In the following, we provide a quick overview of  related contrastive \ac{ssl} techniques and review the attempts made along this line in the field of \ac{mir}, especially for beat and downbeat tracking. We also discuss the recent advances made towards solving these important \ac{mir} tasks.

\subsection{Self-Supervised Representation Learning}

Our approach takes inspiration from contrastive methods.
In CPC (Contrastive Predicting Coding)~\cite{oord2019representation}, representations are learned from sequential data by predicting the future latent representations from the (aggregated) past ones.
For this, an encoder is trained to produce latent representations with the task of making it easy to distinguish in the obtained latent space (positive) future latent representations from a set of negative samples.
This encourages the model to capture meaningful information.
Instead of predicting the future, in Wav2Vec2 \cite{wav2vec2} the task is to predict masked observations.
In Wav2Vec2, features are extracted from an audio signal with a Convolutional Network and fed to a transformer encoder where some frames are masked.
Additionally, the audio features are quantized and the model is trained to contrast the masked output with the quantized output and a set of distractors.

\subsection{\acl{ssl} in \ac{mir}}

Following the trend in speech processing research, \ac{ssl} approaches have started to become popular in \ac{mir}.
On the one hand, these approaches can be used to train general-purpose models, the so-called ``foundation models'' (such as MULE~\cite{mule} or MERT~\cite{DBLP:journals/corr/abs-2306-00107}), which are supposed to be useful to solve a whole set of downstream tasks (see the MARBLE benchmark~\cite{DBLP:conf/nips/YuanMLZCYZLHTDW23}). 
On the other hand, models can be developed to learn representations that are well aligned with a specific \ac{mir} task.
Among those, learning  representations that are equivariant to a semantic distortion of the audio signal has become a popular approach (\textit{e.g.}, for pitch or tempo estimation using siamese networks \cite{spice,quinton,riou2023pesto,antonintempo}).

Few works have proposed to apply \ac{ssl} for rhythm analysis tasks. 
Zero-Note Samba (ZeroNS) \cite{desblancs2023zero} leverages the synchronization of the various instrument stems in a music track. 
For this, they separate music tracks into their percussive and non-percussive parts and train an encoder to force the synchronization between the corresponding latent representations, which are then used for beat tracking.
In \cite{metricalssl} they used binary metric regularity to derive supervision for their CRF loss, enabling the network to model a hierarchical metrical structure.

\subsection{Beat and Downbeat tracking}

Before the rise of deep-learning approaches, beat and downbeat tracking systems were based on two-step systems: first audio features were extracted from the audio signal (including an onset detection function, \acf{plp}, spectral features or a novelty function); then those were used as ``observations'' to a probabilistic model (such as Hidden Markov Models or Dynamic Bayesian Network) \cite{beat_dp,peetersbeat,alonsbeat}.

The shift toward data-driven approaches started with \cite{bocklstm} where the authors proposed to process spectral features with bi-directional Long Short-Term Memory (LSTM) networks.
\cite{bocktcn1} then proposed to replace the LSTM with a \acf{tcn} to process the spectral features.
Later on, the model was improved by solving jointly multiple tasks (beat and downbeat positions, as well as tempo)~\cite{Bck2019MultiTaskLO,Bck2020DeconstructAR}.
Currently, models based on the Transformer architecture, used in a multi-task setting (joint beat-downbeat tracking) are the most successful.
In \cite{Hung2022} the authors apply the Spectral-Temporal Transformer  (SpecTNT) architecture \cite{Lu2021SpecTNTAT} to tackle this task.
This architecture combines a spectral transformer that processes harmonic features and a temporal transformer that aggregates the processed features over time.
To further improve the performance, the authors combined SpecTNT with a \acf{tcn}.
Beat Transformer \cite{ZhaoXW22} incorporates dilated self-attention to capture long-range dependencies.
Furthermore, in the middle layers, they alternate time-wise dilated self-attention with instrument-wise self-attention\footnote{
The instrument-wise attention is conducted along the stems of a demixed audio signal, contributing to a comprehensive analysis of the audio data.}.

\section{Proposed Contrastive Learning SSL scheme}\label{sec:method}

In this paper, we propose a novel \ac{ssl} approach to learn representations useful for rhythm analysis tasks, and instantiate it for the beat tracking downstream task.
We aim to learn a projection (an encoder) such that the resulting projections of observations at  \ac{plp} peaks whose distance from each other is a power of 2 are close to each other, and different otherwise. The two key insights behind this is that: i) a significant fraction of the \ac{plp} peaks  (supposedly aligned to the tatum grid) is expected to represent beat positions, with high probability, and ii) most of the musical recordings tend to have a binary metric structure (i.e. beats can be musically divided by two and grouped by two). We conjecture that despite being over-simplistic, these ingredients are ``good-enough'' to define a pretext-task that will be effective for training representations useful for various rhythm analysis tasks, especially beat tracking, provided that a downstream fine-tuning phase is anyway envisaged. In the following we will refer to the distance between two \ac{plp} peaks as \textit{tatum-unit} and denote it by $tu$.

\begin{figure}[t]
  \centerline{
    \includegraphics[width=1\columnwidth]{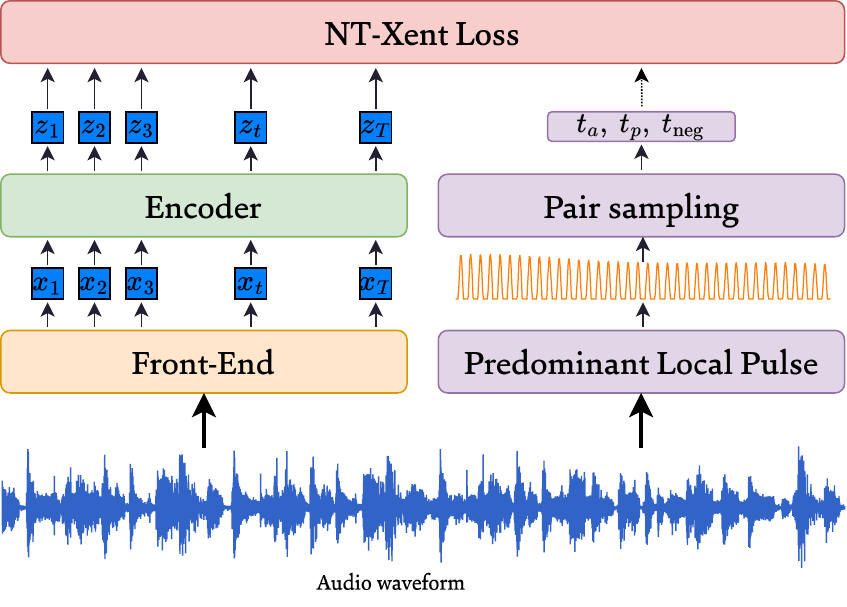}}
  \caption{Our proposed contrastive \ac{ssl} scheme for beat tracking. The left part displays our processed audio waveform to obtain the representations $z_{t}$. The right part displays our mining of positive and negatives.}
  \label{fig:flowchart}
\end{figure}

We solve this pretext-task using contrastive learning.
We learn to distinguish observations at times separated by an interval of a power of 2  in  $tu$  units, from those that are not.
Once computed, the \ac{plp} function is used to select an anchor, its associated positive, and a set of negative samples.
We further explain the procedure in section~\ref{sec:sampling}.
We then train our encoder to attract the anchor and the positive while repelling the set of negatives in the latent space using a contrastive loss.
We describe the architecture of our encoder in section~\ref{sec:model}.
Our approach is summarized in Figure~\ref{fig:flowchart}. 

\subsection{Mining positive and negatives}\label{sec:sampling}

The key part of our work is to learn representations in a contrastive way.
Therefore we need to define an anchor, a positive and multiple negative samples within each given audio excerpt.
We rely on the \acf{plp}~\cite{plp_muller} function  to extract local pulse information (see \ref{subsubsec:plp}).
Given such information, we sample positive and negative times for a selected anchor (see \ref{subsubsec:naive_sampling}).

\subsubsection{Predominant Local Pulse} \label{subsubsec:plp}

The \ac{plp} method analyzes the \ac{osf} of an audio signal in the frequency domain to find a locally stable tempo for each frame.
For this, a ``tempogram'' (a Short-Time Fourier-Transform, STFT) of the \ac{osf} is computed.
At each time position, the maximum of the ``tempogram'' indicates the dominant pulse frequency.
Using the corresponding amplitude and phase of this maximum, one can re-synthesize the corresponding temporal signal (a sinusoidal component).
Using the usual overlap-and-add (OLA) inverse STFT method, a smooth temporal signal is formed by overlapping-and-adding the sinusoidal components with various dominant pulse frequencies over time.
This temporal function is termed \ac{plp} and represents a localized enhancement of the original novelty function's periodicity.

\begin{figure*}[!ht]
  \centerline{
  \includegraphics[width=2\columnwidth]{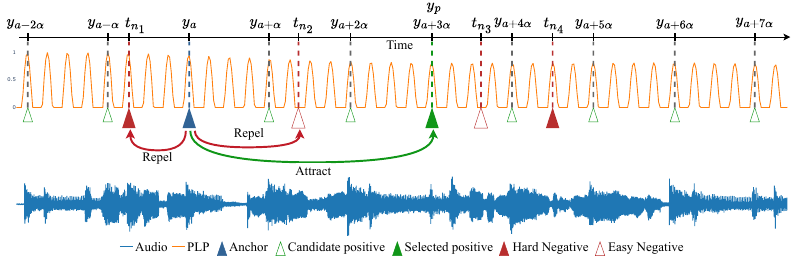}}
  \caption{Proposed mining strategy of Positives and Negatives (easy and hard) given an Anchor time in the \ac{plp} function. Positive are sampled among peaks of the PLP whose time index is distant from the Anchor by a power of two tatum units $tu$ (here $\alpha=4 \times tu$ ); Negatives are the remaining times and are considered Easy if not peaks of the PLP and Hard if peaks of the PLP. Here we sample two hard and two easy negatives.
  }
  \label{fig:pairs_sampling}
\end{figure*}

\noindent\textbf{Computation.} 
Given an audio signal, we compute the \ac{plp} function with the same frame rate as our audio front-end (\textit{i.e} 20ms).
We used the \textit{beat.plp} function from Librosa~\cite{mcfee2015librosa}. 
The function is fed with an \ac{osf} computed from a spectrogram\footnote{In a preliminary experiment, we found that using the spectrogram to get the \ac{osf} was working better than using the Mel-spectrogram} with 2048 points and the default minimum and maximum tempo parameters.
We then estimate the local maxima peak $y_k$ of the \ac{plp} using the \textit{find\_peaks} function from scipy.

\subsubsection{Sampling from PLP}\label{subsubsec:naive_sampling}

In the following we will refer to the distance between two \ac{plp} peaks as \textit{tatum-unit} and denote it by $tu$.
For simplification, we do the following assumptions.
We assume that the tatums correspond to the 8-th note and that most tracks are in a 4/4 meter.\footnote{In a preliminary experiment, we delve deeper into determining the metrical level that the peaks of the \ac{plp} correspond to. 
Our findings suggest that these peaks align with either the beat, the 8-th note or the 16-th note level.}
Following this, we consider that the positives have a time distance $\Delta$ from the anchor which is a power of two of the tatum unit: $\Delta= i \times \alpha \times tu$ with $\alpha = 2^{n}$ and $i \in \mathbb{Z} \setminus \{0\}$.
In this work we consider $n=2$ (which corresponds to an inter-distance of two beats).

We define by $Y = \{y_1, ..., y_K\}$  the set of \ac{plp} peaks within a given audio segment.
We first sample an anchor $a$ uniformly in $[1, K]$.
We denote by $y_a$ the time associated to $a$ (blue arrow in Fig.\ref{fig:pairs_sampling}).
Given this anchor, we sample its associated positive time step $p$.
This positive must be situated $i \times \alpha$ peaks away from the query.
For a given anchor $a$, we therefore sample $p$ uniformly from $Y_{a} = \{y_{a \pm i \times \alpha},  0  \leq a \pm i \times \alpha \leq K   \}$.

We denote by $y_p$ the time associated to $p$ (green arrow in Fig.\ref{fig:pairs_sampling}, green empty arrows are all the elements of $Y_{a}$).

We then sample $N$ negative time steps at which we define hard negative and easy negative examples.
An \textit{easy negative} corresponds to a time step that is not a \ac{plp} peak.
They are sampled uniformly in $[0, T] \setminus Y$.
We also apply a ``safety window'' (whose duration was empirically determined to one frame) around peak time steps to avoid sampling negatives that are too close to a peak.
A \textit{hard negative} corresponds to a time step that is a peak but that is not in $Y_a$.
They are sampled uniformly in $Y  \setminus (Y_a \cup \{y_a\})$.
We sample $N=10$ negatives, half of them are hard negatives, and the other half are easy negatives.

To prevent any errors coming from the \ac{plp} function we discard audio segments where the inter-peak distance is not almost constant.
We empirically set the allowed variation to 20 percent of this inter-peak distance within a segment (more details about this are given on the companion website).

\subsection{Architectures}\label{sec:model}

\paragraph{Front-end.}
We compute Mel spectrogram features from audio sampled at 16kHz using 128 bands, a window size of 2048 samples, and a hop size of 320 samples (20ms frame rate).
We apply log compression and normalization\footnote{
For normalization, we use the mean and standard deviation computed over the training set.}.
Subsequently, a linear layer projects the frames to the embedding dimension.
The resulting sequence $x_t$ serves as the input to the encoder.
We use audio segments of 20s long to ensure the model sees a sufficiently large context.
However, we did not explore varying the length of audio segments fed into the encoder.

\noindent
\textbf{Encoder.}
For the encoder, we use a Transformer architecture similar to the one used in Wav2Vec2 or Hubert~\cite{wav2vec2,Hubert}.
It is composed of a stack of Transformer encoder layers.
Each layer is composed of a multi-head self-attention mechanism followed by a feed-forward network.
We use 8 layers each of which has 8 attention heads and apply a 0.1 dropout in the attention layer.
The encoder outputs the embedding sequence $z_t$.
The embedding dimension is set to 512 and the hidden dimension of the feed-forward network is set to 1024.
In total, the model has 19.1M learnable parameters.
We did not explore other architectures as the focus of our work was to study the proposed \ac{ssl} scheme.

\subsection{Contrastive Loss}
Among the various formulations of the contrastive losses, we have chosen to use the NT-Xent loss~\cite{simclr} one.
We define the similarity measure between two vectors $u$ and $v$ as $\textup{sim}(u,v) = \frac{u^\top v}{\parallel u \parallel \parallel v \parallel}$.
Given an anchor $y_a$, a positive $y_p$ and a set of $N$ negatives time-steps $t_{\text{neg}} = \{t_{n_1}, ..., t_{n_N}\}$,  we compute the contrastive loss as follows:
\vspace{-0.2cm}
\begin{equation}
  \mathcal{L}_{\text{NT-Xent}}(y_a,y_p,t_{\text{neg}}) = - \log \frac{\exp(sim(z_{y_a},z_{y_p})/\tau)}{\sum_{i=1}^{N} \exp(sim(z_{y_a}, z_{t_{n_i}})/\tau)}.
\end{equation}

We set the temperature to $\tau = 0.1$. 
For each audio in a batch, we use 80\% of the available peaks as anchors.
For each of them, we sample their corresponding positive and negatives.
We compute the above contrastive loss over each pair and each audio.
We then average the losses to obtain the global loss for the batch, that is if we have a total of $M$ pairs in the batch:
\vspace{-0.2cm}
\begin{equation}
  \mathcal{L} = \frac{1}{M}\sum_{y_a, y_p,t_{\text{neg}}}^{}\mathcal{L}_{\text{NT-Xent}}(y_a, y_p,t_{\text{neg}}).
\end{equation}

\section{Evaluation}\label{sec:expe}

To evaluate our model, we performed three experiments. 
In all three experiments, the model is pre-trained in a \ac{ssl} way using unlabeled data.

In Experiment 1, we test the \ac{fsl} abilities of our model using only a few data for fine-tuning.
Experiment 2 tests the generalization of our model on unseen conditions and serves as comparison to ZeroNS.
Finally, Experiment 3 compares our performance to the ones obtained using fully-supervised beat-tracking models. 

\subsection{Datasets}

For \ac{ssl} pre-training, we use a combination of unlabeled datasets (in terms of beat positions): the Free Music Archive (FMA) \cite{fma_dataset}, MTG-Jamendo \cite{bogdanov2019mtg}, and MagnaTagaTune (MTT) \cite{Law2009EvaluationOA}.
FMA contains 106,574 full tracks spanning 161 genres.
MTG-Jamendo contains around 55,000 full audio tracks.
Finally, MTT contains approximately 26,000 excerpts of 30-s duration from 5223 unique tracks.
Overall the combined datasets offer around 165k full audio tracks and a total of 8,000 hours.

For fine-tuning and testing, we used the following labeled (into beats) datasets, commonly used in previous works:
SMC \cite{SMC}, Ballroom \cite{ballroom} and Hainsworth \cite{hainsworth}, GTZAN \cite{GTZAN,GTZAN_2}, RWC \cite{Goto2002RWCMD} and  Harmonix \cite{oriol_nieto_2019_3527870}.
The Harmonix dataset is mainly composed of pop music tracks, whereas the Ballroom, GTZAN, RWC, and Hainsworth datasets offer a wider variety of musical genres.

\subsection{Evaluation Metrics}
We report the commonly used metrics in the literature including the F-measure with a tolerance window of $\pm$ 70ms, continuity-based measures at the correct metrical level (CMLt \& CMLc), and at alternate metrical levels such as double/half and offbeat (AMLt \& AMLc) \cite{evaluation_beat}.

\subsection{Implementation details}

\subsubsection{Pre-training}
For \ac{ssl} pre-training we kept 0.05\% of the data for validation (9,000 tracks).
Our model is pre-trained during 200 epochs (equivalent to around 270,000 steps).
Training was conducted on 4 A100 GPUs utilizing float 16 precision and a global batch size of 96.
We employed the Adam optimizer \cite{adam} with an initial learning rate set at 1e-4 and applied a polynomial decay learning rate scheduler.
The learning rate gradually increased to 5e-4 within the first 32,000 steps, then reverted to its initial value over the subsequent 250k steps.
Additionally, gradient clipping was employed.
We keep the model that gives the best validation loss.

\subsubsection{Fine-tuning}\label{sec:finetuning}

After \ac{ssl} pre-training, we need to adapt the model to the downstream task of beat tracking.
This is done by adding a linear classification probe $g(.)$ and fine-tuning both the encoder and the linear probe.
$g(.)$ projects the embedding into the scalar beat activation function.
Instead of feeding $g(.)$ with the output of the encoder, we feed it with a weighted sum of the outputs of each layer of the Transformer~\cite{Yang2021SUPERBSP}.
That is $z = \sum_{l=1}^{8} \alpha_l z^{(l)}$, where $z^{(l)}$ is the output of layer $l$.
The weights $\alpha_l$ are jointly learned with the linear probe $g(.)$.

The system is trained to minimize the binary cross-entropy loss between the beat activations and the target.
Following the literature we widened the beat targets by a window [0.25, 0.5, 1, 0.5,0.25] \cite{Bck2020DeconstructAR}.
We used the Adam optimizer \cite{adam} with an initial learning rate of 1e-5 and a polynomial decay learning rate scheduler.

During fine-tuning, we utilized audio chunks of sizes similar  to those used during pre-training (20s).
However, during inference, to avoid potential out-of-memory errors, we split audio excerpts exceeding 45 seconds into 20-second chunks with 5-second overlap.
Subsequently, we overlap-add the activations to derive the beat activations for the whole track.
These beat activations are then fed into a \acf{dbn} \cite{florian_krebs_2018_1414966} to predict the beat positions.
The \ac{dbn} is configured to model a tempo range of 40-270  beats per minute with transition lambda set to 45, observation lambda to 9, and a threshold of 0.15.

\subsubsection{Data Augmentation}

We found that both pre-training and fine-tuning could benefit from data augmentation, in particular time-stretching.
We apply time-stretching in two manners: constant factor  and time-varying factor.
In both cases, we constrain the time-stretching factor to lie in the interval $[0.8, 1.2]$.  
For the constant factor case, we used sox effects in Torchaudio \cite{hwang2023torchaudio}, and for time-varying factor we used  Libtsm \cite{libtsm}.
When using a time-varying factor we randomly sample time instants at which the stretching factor is modified (also randomly, see the repository for details).
This was found to be particularly beneficial for the pre-training stage.
Indeed because we have filtered out tracks where the inter-peak distance is not almost constant, the \ac{ssl} training data does not contain examples of time-varying tempo.
Using time-varying time-stretching allows us to simulate this in a controlled fashion.

We found that it was better to compute the \acf{plp} curve before time-stretching and shift the peaks accordingly, rather than on the time-stretched audio.

\subsection{Experiment 1: Few-shot learning}

\begin{figure*}[!ht]
  \centerline{
    \includegraphics[width=1.9\columnwidth]{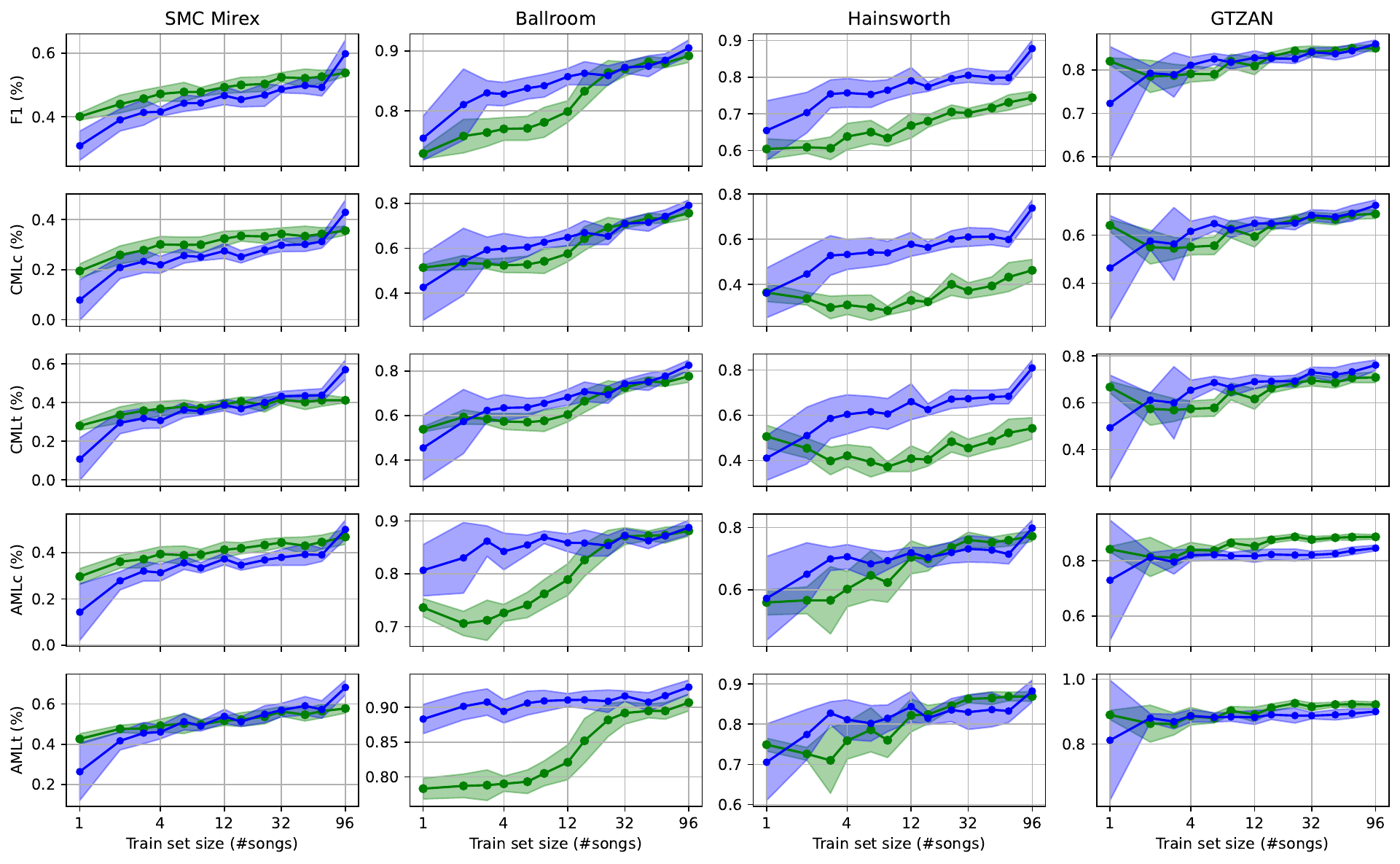}}
  \caption{Results of Experiment 1: Few-Shot Learning. Shaded areas representation the standard deviation. (ZeroNS in green and our method in blue)}
  \label{fig:few_shot}
\end{figure*}

\textbf{Protocol. }
The goal here is to test the ability of our model to learn with only few examples, \acf{fsl}.
To be able to compare our results with previously published ones, we replicate the evaluation protocol proposed in ZeroNS~\cite{desblancs2023zero}.
We consider \textit{individually} each dataset (both for fine-tuning and testing): $T \in $ \{SMC, Ballroom, Hainsworth, GTZAN\}.
For each dataset $T$, we split it into 8 folds, we use one for testing $T_{test}$, one for validation $T_{valid}$ and perform \ac{fsl} with the remaining ones $T_{train}$.
The \ac{fsl} ability is evaluated by selecting randomly $k \in \{1,2,3,4,6,8,12,16,24,48,64,96\}$ items from $T_{train}$.
For each $k$ we sample 10 variations: $T_{train,i}^k$
For each choice of $k$, we fine-tune our pre-trained model on each variation $T_{train,i}^k$ and keep the one that performs the best on $T_{valid}$

\noindent\textbf{Results.}
We give the results in Figure~\ref{fig:few_shot} for the $T_{test}$ of each dataset (SMC Mirex, Ballroom, Hainsworth, and GTZAN) and each value of $k$ (x-axis).
We report the mean and standard deviation of the metrics over the training set variations $T_{train,i}^k$.
Our model performs at least as well as ZeroNS on almost all metrics and datasets.
The exceptions are with AMLt on SMC and AMLc and AMLt on GTZAN and SMC.
We observe that our model performs significantly better on Hainsworth, with up to 10\% absolute improvement in F1 score and almost 20\% absolute improvement in CMLt and CMLc.
Also, the performance gap is significant on Ballroom when using very few data (less than 10 tracks) where we can observe almost 10\% absolute improvement in F1 score and up to 15\% improvement in AMLc.

\subsection{Experiment 2: Generalization}

\textbf{Protocol.}
The goal here is to test the generalization ability of our model, \textit{i.e.} training our model on one dataset and testing on another.
For this, we replicate the protocol proposed in ZeroNS~\cite{desblancs2023zero}.
For each choice of dataset $T \in$ \{SMC, Hainsworth, Ballroom\}, we split it into 8 folds, we use one for validation $T_{valid}$, and the remaining seven for training $T_{train}$.
We then use the best-performing model on $T_{valid}$.
Instead of using the linear probe described above, we obtained better results using a MLP (two linear layers interleaved with a ReLU), also fed by the weighted sum of layer sequences (sec \ref{sec:finetuning}).
Whatever the choice of $T$, the test is performed on the GTZAN dataset.
\begin{table}[!ht]
  \scriptsize
  \begin{tabular}{llcccc}
    \toprule
    \textbf{Trained on}         & \textbf{Method} & \textbf{F1} (\%)        & \textbf{AMLt} (\%)      & \textbf{CMLt}  (\%)     \\
    \midrule
    \multirow{2}{*}{SMC}        & Ours            & $\mathbf{79.5 \pm 0.5}$ & $\mathbf{88.0 \pm 0.6}$ & $\mathbf{64.4 \pm 0.9}$ \\
                                & ZeroNS          & $74.8 \pm 2.1$          & $86.3 \pm 2.3$          & $51.0 \pm 2.1$          \\
    \addlinespace
    \multirow{2}{*}{Hainsworth} & Ours            & $\mathbf{85.1 \pm 0.8}$ & $\mathbf{89.9 \pm 0.9}$ & $\mathbf{73.2 \pm 1.8}$ \\
                                & ZeroNS          & $80.6 \pm 0.9$          & $89.4 \pm 0.7$          & $62.8 \pm 2.3$          \\
    \addlinespace
    \multirow{2}{*}{Ballroom}   & Ours            & $\mathbf{83.9 \pm 0.3}$ & $88.4 \pm 0.5$          & $\mathbf{72.3 \pm 0.9}$ \\
                                & ZeroNS          & $82.6 \pm 0.5$          & $\mathbf{89.0 \pm 0.8}$ & $67.6 \pm 1.1$          \\
    \bottomrule
  \end{tabular}
  \caption{Results of Experiment 2: Generalization}
  \label{tab:gtzan_gen}
\end{table}

\noindent\textbf{Results.}
We indicate the results in Table. \ref{tab:gtzan_gen}.
We report the mean and standard deviation of F1, AMLt, and CMLt scores across the different folds.
Overall our model performs better than ZeroNS on all datasets except when for the AMLt metric when trained with Ballroom, but the difference is not statistically significant.
This means that our model can generalize well to unseen data.
Precisely we observe a 5\% improvement in F1 score and more than 10\% improvement in CMLt when training on SMC or Hainsworth.
We nearly reach the F1 score of fully supervised models (presented next) when training solely on ${7/8}$ of Hainsworth (\textit{i.e} 194 tracks).

\begin{table}[t]
  \centering
  \scriptsize
  \resizebox{0.7\columnwidth}{!}{
    \begin{tabular}{l|ccc}
      \toprule
      Method                           & F1            & CMLt          & AMLt \\
      \midrule
      Böck \cite{Bck2020DeconstructAR} & 0.885          & \textbf{0.813}  & \textbf{0.931}     \\
      Hung \cite{Hung2022}             & \textbf{0.887}  & 0.812  & 0.920       \\
      Zhao \cite{ZhaoXW22}             & 0.885          & 0.800   & 0.922        \\
      Ours                             & 0.876           & 0.802   & 0.918        \\
      \bottomrule
    \end{tabular}%
  }
  \caption{Results of Experiment 3: Comparison with supervised baseline}
  \label{tab:cross_val}
\end{table}

\subsection{Experiment 3: Comparison with supervised baseline}
\textbf{Protocol.}
The goal here is to compare the performance of our model to the ones provided by fully-supervised models.
For this we replicate the commonly used 8-fold cross validation set-up after \cite{Bck2020DeconstructAR,Hung2022,ZhaoXW22}.
GTZAN is kept as a test set and is never seen in training.
We average the metrics over the 8 training folds to obtain the final results.

\noindent\textbf{Results.}
We give the results in Table \ref{tab:cross_val}.
It is clear that the proposed beat tracking approach  using our  self-supervised pre-training can be competitive with state-of-the-art methods on GTZAN, a dataset covering a wide diversity of genres.
While our method does not outperform the best-performing method, it achieves comparable results across all metrics, proving the quality of the learned representations.

\vspace{-1em}
\section{Conclusion}\label{sec:conclusion}

In this paper, we proposed a novel \acl{ssl} approach to learn representations useful for the task of beat tracking using contrastive learning where the selection of anchor, positive and negative peaks derives from a \acl{plp} function.

We assess our proposal positively based on a series of experiments.
In a first experiment, we showed that our proposed approach was superior on some datasets to the previous SSL approach, ZeroNS, in a few-shot learning setting.
In a second experiment, we show that our model has better generalization capabilities to unseen data.
In the last experiment, we show that our model also yields comparable performances to the fully supervised baseline, indicating that our pre-training scheme effectively learns meaningful beat-related representations.

To further improve our method, future work will focus on developing a more sophisticated sampling mechanism that can handle other metrical structures than the binary one used-here (such as 6/8, 3/4).
One potential approach is to incorporate additional audio features, such as self-similarity matrices, to gain a deeper understanding of the rhythmic structure within an audio segment and adaptively select positive positions for a given anchor.

\section{Acknowledgements}

This work was granted access to the HPC resources of IDRIS under the allocation 2022-AD011013924R1 made by GENCI. The material contained in this document is based upon work funded by the ANR-IA and Hi! PARIS.

\bibliography{ISMIRtemplate}

\end{document}